\documentclass[11pt]{article}

\usepackage[tbtags]{amsmath}
\usepackage{graphicx}
\begin{document}
\topmargin 0pt
\oddsidemargin 0mm
\renewcommand{\thefootnote}{\fnsymbol{footnote}}
\begin{titlepage}

\vspace{5mm}

\begin{center}
{\Large \bf Effect of a relativistic correction to the Coulomb potential on the energy levels of hydrogen atom} \\

\vspace{6mm} {\large Harihar Behera\footnote{E-mail: harihar@iitb.ac.in;  
 harihar@iopb.res.in }}   \\
\vspace{5mm}
{\em
Department of Physics, Indian Institute of Technology, Powai, Mumbai-400076, India} \\

\end{center}

\vspace{5mm}
\centerline{\bf {Abstract}}
\vspace{5mm}
 Based on classical electrodynamics, it is argued that the Coulomb potential (which is strictly valid for two point charges at rest), commonly used in the study of energy levels of hydrogen atom is not the correct  one, because the electron in the hydrogen atom moves with relativistic speeds with respect to the nucleus. Retardation effect has to be considered in accordance with  Li\'{e}nard-Wiechert (or retarded) potential of a moving charge or the relativistic electrodynamics. However, such a consideration introduces a correction to the Coulomb potential, whose quantum mechanical expectation value is estimated at $E_{ret} = - \frac{mc^2\alpha ^4}{2n^3(l+1/2)}$, which is of the same order as the fine structure of hydrogen atom and hence added to the standard energy eigenvalue values of H-atom. This correction lifts the $l$-degeneracy in the spectra of H-atom and hence modifies the standard result. The result disturbs the existing agreement between the theory and experiments on H-atom and hence requires further theoretical and experimental re-examination. The implications of this result for the Kepler-problem in general is also discussed in the context of Heaviside's gravity, which seems to offer an alternative explanation for the non-Newtonian perihelion advance of Mercury without 
 invoking the space-time curvature formalism of Einstein's general theory of relativity.  \\


\end{titlepage}
In both non-relativistic quantum theory \cite{1,2} and Dirac's relativistic quantum theory \cite{3} of hydrogen atom one considers the potential energy of a hydrogen atom as the Coulomb potential energy:
\begin{equation}
V_C(R)= \frac{1}{4\pi \epsilon _0}\frac{Qq}{R}
\end{equation}
 where $Q = e$ is the electric charge of the nucleus of H-atom and $q =-e $ is 
electric charge of electron and $R$ is the instantaneous distance between the point charges $Q$ and $q$. Strictly speaking Coulomb's law of interaction between two point charges is valid for the stationary charges. In the H-atom problem, one of the charges, viz., the electron, moves around the heavy nucleus (assumed stationary) at relativistic speeds, so the interaction between the electron and the nucleus is not strictly Coulombic (which is valid for charges at rest). Relativistic modification to $V_C(R)$ is required for an additional correction to the energy levels of H-atom, which we report here. From classical electrodynamics [4,5], we know that the scalar potential of a point charge 
$q$ moving with constant velocity ${\bf v}$ is expressed as \cite{4,5} 
\begin{equation}
V({\bf r}, t) = \frac{1}{4\pi \epsilon _0}\frac{q}{R(1 - v^2sin^2\theta /c^2)^{1/2}} 
\end{equation} 
where ${\bf R} \equiv {\bf r} - {\bf v}t$ is the vector from the {\it present } 
position of the particle to the field point ${\bf r}$, and $\theta$ is the angle 
between ${\bf R}$ and ${\bf v}$ (Fig. 1); $|{\bf v}| = v$, $c$ is the speed 
of light in vacuum and $t$ is the present time. Evidently for non-relativistic velocities 
$(v^2 \ll c^2)$ or for $c = \infty$ ($\Longrightarrow$ action at a distance),
\begin{equation}
V({\bf r}, t) = \frac{1}{4\pi \epsilon _0}\frac{q}{R} \>\>\>\>\mbox{(Coulomb potential)}
\end{equation}
The expression in Eq. (2) is also known as the Li\'{e}nard-Wiechert scalar potential for a point charge moving at constant velocity and there are many ways to obtain it [see page 433 in \cite{5}]. The Li\'{e}nard-Wiechert scalar potential is also known as the retarded scalar potential \cite{4,5}, because if a charge moves in an arbitrary way, the electric potential (or the field) we would find {\it now} at some point depends only on the position and the motion of the charge {\it not now}, but at an earlier time - at an instant which is earlier by the time it would take light going at the speed $c$ to 
travel the distance ${r}$ from the charge to the field point. If another charge Q is at rest at the field point described in Eq. (2) as in Fig. 1, then the interaction between the static charge $Q$ and the moving charge $q$ is expressed as 
\begin{figure}
 \centering
 \includegraphics[scale=1]{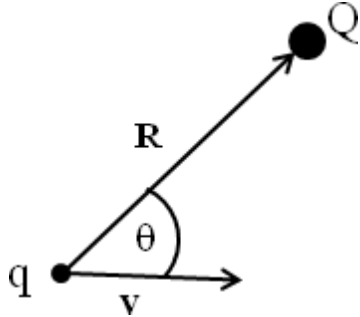}
 \caption{Interaction between a moving charge q and stationary charge Q.}
\end{figure}
\begin{equation}
V({\bf r}, t) = \frac{1}{4\pi \epsilon _0}\frac{Qq}{R(1 - v^2sin^2\theta /c^2)^{1/2}} 
\end{equation}           
which again for $(v^2 \ll c^2)$ or $c = \infty$ ($\Longrightarrow$ action at a distance), reduces to 
\begin{equation}
V({\bf r}, t) = \frac{1}{4\pi \epsilon _0}\frac{Qq}{R} = V_C(R)
\end{equation} 
If $v$ in Eq. (4) represents the instantaneous velocity of the electron with charge $q = -e$ and $Q=e$ represents the charge of an assumed stationary nucleus of an H-atom, 
then Eq. (4) should represent the real potential energy of an H-atom as 
\begin{equation}
V({\bf r}, t) = -\frac{1}{4\pi \epsilon _0}\frac{e^2}{R(1 - v^2sin^2\theta /c^2)^{1/2}} 
\end{equation}
whose effect on the energy levels of an H-atom is considered as follows. Now introducing the angular momentum $L$ of the electron around the nucleus as
$L=|{\bf L}| = |{\bf R}\times (m{\bf v)}|= mvRsin\theta$, where $m$ is the mass of the electron, we rewrite Eq. (6) as
 \begin{equation}
V({\bf r}, t) = -\frac{1}{4\pi \epsilon _0}\frac{e^2}{R(1 - L^2 /(m^2c^2R^2))^{1/2}} 
\end{equation} 
For $L^2 \ll m^2c^2R^2$ and to the first order in $c^2$, we can approximate Eq. (7) as 
\begin{equation}
V({\bf r}, t) \simeq -\frac{e^2}{4\pi \epsilon _0 R}\left(1 + \frac{L^2}{2m^2c^2R^2}+ \ldots\right) 
\end{equation} 
This introduces a $\frac{1}{R^3}$ correction to the Coulomb potential (1) as 
\begin{equation}
H_{ret} = - \frac{e^2L^2}{8\pi \epsilon _0 m^2c^2R^3} 
\end{equation}
\indent
An analogous situation is expected in the Kepler-problem of planetary motion, if one 
considers Heaviside's gravitational field equations \cite{7,8,9,10,11}, which are analogous to the Maxwell's equations of electromagnetism (with the differences that the source terms in Heaviside's field equations have signs opposite to that we find in Maxwell's equations of electromagnetism, because of the peculiar attractive nature of gravity between two masses) but less known and recently studied within the framework of special relativity and Dirac's the relativistic quantum mechanics of spin half particles \cite{7} . In Heaviside's gravitational theory, if one takes the 
speed of gravity as $c_g$ \cite{7, 8} (the exact value of which is still unknown experimentally), then a correction to the Newtonian potential in the Kepler-problem can be obtained from Eq.(8), by replacing $e^2$ by $M_\odot m$ ($M_\odot$ is the mass of the Sun, $m$ mass of the planet in question) and $1/4\pi \epsilon _0$ by Newton's universal gravitational constant $G$:
\begin{equation}
H_{ret}^{gravity} = - \frac{GM_\odot L^2}{2mc_g^2R^3} 
\end{equation}
With $c_g = c$ (as per the special relativistic investigation and formulation \cite {7} of Heaviside's gravity), Eq. (10) alone\footnote{Note that the consideration of retarded interactions is most important in planetary motion or other astronomical and cosmological problems. For example, in the case of the Sun and Earth system, any event causing a change in the mass of the Sun would be felt by a change in the motion of the Earth only after 8 minutes and 4 seconds (if $c_g = c$) has elapsed since the occurrence of that event, since light originating from the Sun takes that much of time to reach the Earth.} can predict exactly half the value of the non-Newtonian perihelion advance of a planet such as the Mercury \cite {12, 13, 14} as predicted by Einstein's general theory of relativity \cite{14} and observed experimentally \cite{12,13,14}. In this context, we would like to cite an interesting interpretation of the general relativistic interpretation of the perihelion advance of Mercury as given in a foot-note in page 113 of Thornton-Marion's book \cite{12}:
\begin{quote}
``{\it One half of the relativistic term results from effects understandable in terms of 
special relativity, viz., time dilation $(1/3)$ and the relativistic momentum 
effect $(1/6)$; the velocity is greatest at the perihelion and least at aphelion 
(see Chapter 14). The other half of the term arises from general relativistic 
effects and is associated with the finite propagation time of gravitational interactions. Thus, the agreement between theory and experiment confirms the 
prediction that the gravitational propagation velocity is the same as that for light.}"
\end{quote}   
Thus, it seems one can explain the non-Newtonian perihelion advance of Mercury 
with the help of Heaviside's gravity and special relativity without invoking 
the space-time curvature formalism of general relativity. Now let us not 
discuss this gravitational problem further and come back to the present 
problem of H-atom.\\ 
\indent
We can calculate the effect of the correction Eq. (9) to the Coulomb potential on the energy levels of an H-atom by treating this as a correction to the non-relativistic Hamiltonian  
\begin{equation}
H = - \frac{\hbar ^2}{2m}\nabla ^2 - \frac{e^2}{4\pi \epsilon _0}\frac{1}{R} 
\end{equation}
For this purpose, we will now consider the classical variables 
$L^2$ and $\frac{1}{R^3}$ as quantum mechanical operators according to the postulates of
quantum mechanics. Considering Eq. (9) as a perturbation potential and using the 
time-independent 1st order perturbation theory \cite{1,2} with $\Psi _{nlm}$ as the unperturbed state, the expectation value of $H_{ret}$ was then calculated to give an energy 
eigenvalue 
\begin{equation}
E_{ret} = \left<H_{ret}\right> = -\frac{1}{2}mc^2\frac{\alpha ^4}{n^3(l+\frac{1}{2})}
\end{equation}
which is of the same order of magnitude as the fine-structure of H-atom, where 
$\alpha = e^2/(4\pi \epsilon _0 \hbar c)= 1/137.0363$ is the fine-structure constant and other 
symbols have their usual meanings. 
Combining this with the Bohr energy obtainable from Eq.(11), the fine-structure energy \cite{1,2}(due to relativistic correction to kinetic energy of the electron and spin-orbit interaction) we get a modified result for the energy levels of H-atom:
\begin{equation}
E_{njl} = -\frac{1}{2}mc^2\frac{\alpha ^2}{n^2}\left[1 + \frac{\alpha ^2}{n^2}\left(\frac{n}{(j + 1/2)} + \frac{n}{(l + 1/2)} - \frac{3}{4}\right)\right]  
\end{equation}
 The degeneracy in $l$ is not broken in Eq. (13), in contrast with standard energy levels \cite{1,2,3} (as in the Dirac theory of H-atom), viz.,
\begin{equation} 
 E_{nj} = -\frac{1}{2}mc^2\frac{\alpha ^2}{n^2}\left[1 + \frac{\alpha ^2}{n^2}\left(\frac{n}{(j + 1/2)} - \frac{3}{4}\right)\right]  
\end{equation}  
where the breaking of the $l$-degeneracy is evident. For instance, the states $^2P_{1/2}$ and $^2S_{1/2}$ which were degenerate according to Eq. (14) are now non-degenerate by Eq.(13). It is to be noted that Lamb shift introduces a tiny splitting of the $^2P_{1/2}$ and $^2S_{1/2}$ levels, which is of the order of $mc^2\alpha ^5$ \cite{1,2}. In the present correction, the 
splitting of these two levels is of the order of $mc^2\alpha ^4$ and energy level shift is opposite to that of the Lamb shift \cite{1,2} arising out of the quantization of the coulomb field. Thus, a serious disagreement between the theory and experiments on H-atom arises, if we take into account the present correction to the hydrogen energy levels, which needs further theoretical and experimental re-examination of this study.\\  
Further, if, as in the relativistic Kepler-problem of planetary precession, relativistic 
time-dilation is to be considered, then an additional correction term would 
be $E_t = (2/3)E_{ret}$. On addition of $E_t$ to Eq. (13), we get
\begin{equation}
E_{njl} = -\frac{1}{2}mc^2\frac{\alpha ^2}{n^2}\left[1 + \frac{\alpha ^2}{n^2}\left(\frac{n}{(j + 1/2)} + \frac{5n}{3(l + 1/2)} - \frac{3}{4}\right)\right]  
\end{equation}
This further amplifies the disagreement between the theory and experiments. Since the points raised here are of fundamental importance and the predicted results are disturbing the established results, we invite  further theoretical as well as experimental re-examination or scrutiny of our results for a resolution the problem  raised here.


\begin{thebibliography}{99}
\bibitem{1} D. Griffiths, {\it Introduction to Quantum Mechanics}, (Prentice Hall, Inc. NJ, 1995).  
\bibitem{2} S. Gasiorowicz, {\it Quantum Physics}, 3rd Ed.(John Wiley \& Sons, Inc., NJ, 2003). 
\bibitem{3} W. Greiner, {\it Relativistic Quantum Mechanics}, 3rd Ed. (Springer-Verlag, Berlin, 2000). 
\bibitem{4} J. R. Reitz, F. J. Milford, and R. W. Christy, {\it Foundations of Electromagnetic Theory}, 3rd Ed. (Reading, MA: Addison-Weley, 1979). 
\bibitem{5} D. J. Griffiths, {\it Introduction to Electrodynamics}, 3rd Ed. 
(Prentice Hall, Inc. New Jersey, 1999). 
\bibitem{6} O. Heaviside, A gravitational and electromagnetic Analogy, Part I,
The Electrician, 31, 281-282 (1893); Part II, The Electrician, 31, 359 (1893). 
Other references and recent developments of Heaviside's work can be seen in \cite{7,8,9,10,11}
\bibitem{7} H. Behera, P. C. Naik, Int. J. Mod. Phys. A, 19, 4207 (2004). The relativistic version of Heaviside's gravity is what these authors call Maxwellian gravity for historical reasons, since J. C. Maxwell had made the first attempt for such a gravitational theory but did not proceed further for some issues discussed and addressed in \cite{7}. 
\bibitem{8} H. Behera, Newtonian Gravitomagnetism and analysis of Earth Satellite Results, arXiv:gr-qc/0510003.
\bibitem{9} O. D. Jefimenko, {\it Causality electromagnetic induction and 
gravitation: A different approach to the theory of electromagnetic and gravitational fields}, 2nd Ed. (Electret Scientific Company, Star City, 2000); Appendix 8 contains an
unedited copy of Heaviside's original work except that some formulas and all vector equations have been converted to modern notation. Jefimenko's derivation and discussions 
of Heaviside's gravitational field equations are based on the principle of causality. 
\bibitem{10} L. Brillouin, {\it Relativity Reexamined},(Academic Press, New York, 1970); pp-101-104. 
\bibitem{11} K. T. McDonald, Am. J. Phys. 65, 591 (1997). 
 \bibitem{12} S. T. Thornton, J. B. Marion, {\it Classical Dynamics of Particles and Systems}, 5th Ed. (Thomson Books/Cole, Singapore, 2004), 313. 
\bibitem{13} H. Goldstein, {\it Classical Mechanics}, 2nd Ed. (Addison Wesley Pub. Co. Inc., Massachusetts, 1980).
\bibitem{14} S. Weinberg, {\it Gravitation and Cosmology: Principles and Applications of the General Theory of Relativity}, (John Wiley \& Sons, Inc. New York, 1972). 








\end{thebibliography}
\end{document}